\documentclass[11pt,oneside,letterpaper]{article}
\usepackage{amssymb}
\usepackage{amsmath}
\usepackage[dvips]{graphicx}
\usepackage{setspace}
\usepackage{fancyhdr}
\usepackage{ifpdf}
\usepackage{graphicx}
\usepackage{comment}

\newcommand{\bea}{\begin{eqnarray}}
\newcommand{\eea}{\end{eqnarray}}
\newcommand{\be}{\begin{equation}}
\newcommand{\ee}{\end{equation}}

\newcommand{\bi}{\begin{itemize}}
\newcommand{\ei}{\end{itemize}}

\numberwithin{equation}{section}
\allowdisplaybreaks  

\begin{document}

\onehalfspacing

\begin{center}
{ \LARGE \textsc{{{A De Sitter Hoedown\footnote{
Hoedown: American folk dance characterized by rotation of partners \cite{hoedown}.
}}}}\\}
\begin{center}
Dionysios Anninos and Tarek Anous
\end{center}
\end{center}
\vspace*{0.6cm}
\begin{center}
Center for the Fundamental Laws of Nature\\
Jefferson Physical Laboratory, Harvard University, Cambridge, MA 02138, USA
\vspace*{0.8cm}
\end{center}
\vspace*{1.5cm}
\begin{abstract}
Rotating black holes in de Sitter space are known to have interesting limits where the temperatures of the black hole and cosmological horizon are equal. We give a complete description of the thermal phase structure of all allowed rotating black hole configurations. Only one configuration, the rotating Nariai limit, has the black hole and cosmological horizons both in thermal and rotational equilibrium, in that both the temperatures and angular velocities of the two horizons coincide. The thermal evolution of the spacetime is shown to lead to the pure de Sitter spacetime, which is the most entropic configuration. We then provide a comprehensive study of the wave equation for a massless scalar in the rotating Nariai geometry. The absorption cross section at the black hole horizon is computed and a condition is found for when the scattering becomes superradiant. The boundary-to-boundary correlators at finite temperature are computed at future infinity. The quasinormal modes are obtained in explicit form. Finally, we obtain an expression for the expectation value of the number of particles produced at future infinity starting from a vacuum state with no incoming particles at past infinity. Some of our results are used to provide further evidence for a recent holographic proposal between the rotating Nariai geometry and a two-dimensional conformal field theory.

\end{abstract}

\newpage
\setcounter{page}{1}
\pagenumbering{arabic}

\section{Introduction}

Quantum gravity in a de Sitter background remains a beautiful puzzle in theoretical physics \cite{Hawking:2000da,Banados:1998tb,Banks:2003cg,Banks:2005bm,Banks:2006rx,Witten:2001kn,Strominger:2001pn,Maldacena:2002vr,Bousso:1998bn,Polyakov:2007mm,Tsamis:1996qm}. Due to the presence of a cosmological horizon attached to any given observer it contains, according to Gibbons and Hawking \cite{Gibbons:1977mu}, an entropy given by the area of the horizon. However, unlike a black hole horizon, this horizon can never be reached or probed by a local observer. Furthermore, no observer can access the global geometry of de Sitter space. The boundaries of global de Sitter space are the spacelike hypersurfaces given at future and past infinity $\mathcal{I}^\pm$, where it has been suggested that a putative holographic theory might reside \cite{Witten:2001kn,Strominger:2001pn}.

An interesting part of the classical de Sitter spectrum are the de Sitter black hole geometries. In a theory of pure gravity endowed with a positive cosmological constant, the most general known black hole solution is the Kerr-de Sitter black hole. Unlike black holes in asymptotically flat or anti-de Sitter space, de Sitter black holes are bounded in size by the cosmological horizon. The limiting case is given by the rotating Nariai solution \cite{Booth:1998gf,nariai}, which is the near horizon geometry between the black hole and cosmological horizons in the limit where they coincide. Geometrically, it is given by an $S^2$ fibration over two-dimensional de Sitter space.

We begin our explorative work by examining the thermodynamics of general Kerr-de Sitter black holes \cite{Davies:1989ey,Dehghani:2002np}. We treat the black hole and cosmological horizons as thermal entities in their own right and obtain the regions of phase space where they have positive and negative specific heats. Generally, the black hole horizons are out of thermal equilibrium with the cosmological horizon. However, there are three limits where one can define a Euclidean instanton associated to the Lorentzian spacetime \cite{Booth:1998gf}. Firstly, the black hole may be extremal, in which case the Euclidean time coordinate need not be periodically identified. Secondly, there is the lukewarm solution which is defined by the black hole and cosmological horizons sharing the same temperature. Finally, there is the rotating Nariai geometry where the black hole and cosmological horizons approach each other.

Of the three limits, it is only in the rotating Nariai geometry that the angular velocities of the black hole and cosmological horizons tend to coincide. Thus, even though the lukewarm configuration may be in thermal equilibrium, it is out of rotational equilibrium and will generally exchange particles carrying angular momentum, a process enhanced by superradiance at the quantum level \cite{Tachizawa:1992ue}. We will find that this effect is absent in the rotating Nariai limit. Upon perturbations of this spacetime, however, thermodynamic evolution of the system leads it to the most entropic configuration - pure de Sitter space. Thus, the equilibrium of the rotating Nariai geometry is found to be unstable. Even so, this geometry is interesting in its own right given that it is mediated from a Euclidean instanton and can thus serve as a natural starting point in the thermal evolution.

In the second part of the paper we focus on the rotating Nariai geometry and in particular we consider massless scalar waves about this geometry. We find explicit solutions to the wave equation, which are given explicitly by hypergeometric functions. Equipped with these solutions we proceed to compute the quasi-normal modes of the rotating Nariai geometry by imposing that the waves are purely ingoing at the black hole horizon and purely outgoing at the cosmological horizon (see for example \cite{Brady:1999wd}). These quasinormal modes encode the dissipative information of the spacetime upon scalar perturbations. We find two quantization conditions, one related to the frequency and the other to the axial angular momentum of the modes.

It is also of interest to compute the boundary-to-boundary correlators at $\mathcal{I}^+$ in the thermal background given the possible holographic relevance of the $\mathcal{I}^+$ boundary \cite{Anninos:2009yc}. The basic structure of the correlators is fixed by the boundary condition forcing the modes to be purely ingoing at the black hole horizon. The thermal boundary-to-boundary correlators are thus computed by taking variational derivatives of the matter action with respect to the boundary value of the scalars. They take precisely the form of a two-point function in a two-dimensional conformal field theory at finite temperatures as we shall soon discuss.

Subsequently, we compute the absorption cross-section $\sigma_{abs}$ of the black hole horizon when hit by a beam of particles coming from a region near the cosmological horizon. It is found that the $\sigma_{abs}$ becomes negative for certain values of the frequency, signalling superradiant scattering. Specifically, we find that modes with axial angular momentum $m$ exhibit superradiance when the frequency $\omega$ is within the range, $m\Omega_c < \omega < m\Omega_H$, where $\Omega_H$ and $\Omega_c$ are the angular velocities of the black hole and cosmological horizons. This is in accordance with the result in the full rotating geometry \cite{Tachizawa:1992ue}. The upper bound is expected from a general thermodynamic argument, whereas the lower bound is related to the fact that our incoming wave originates near the cosmological horizon and has to penetrate a barrier near that region.

We then proceed to compute the cosmological particle production probability. In order to do so, the scalar wave equation must be solved in the global patch. Furthermore, one must define $|\text{in}\rangle$ and $|\text{out}\rangle$ vacua as those vacua which are annihilated by positive frequency modes at past infinity $\mathcal{I}^-$ and future infinity $\mathcal{I}^+$ respectively. We find that purely incoming positive frequency modes at $\mathcal{I}^-$ do not annihilate the $|\text{out}\rangle$ vacuum at $\mathcal{I}^+$. Thus there is a non-trivial Bogoliubov transformation between the in- and out-modes and we can explicitly compute the expectation value for the number of particles produced. Furthermore, it is well known that scalar fields in de Sitter space have a complex parameter worth of vacua, the $\alpha$-vacua \cite{Chernikov:1968zm,Mottola:1984ar,Allen:1985ux}. Of these vacua only the Euclidean vacuum, whose positive frequency modes are analytic in the lower hemisphere of Euclidean de Sitter space, reduces to the Minkowski vacuum at arbitrarily short distances. Motivated by the importance of the Euclidean vacuum, we propose a definition for positive frequency Euclidean modes in global rotating Nariai as those which are regular in the lower hemisphere of Euclidean $dS_2$, i.e. $S^2$.

Finally, we discuss our results in light of the proposal that the rotating Nariai geometry is holographically dual to a two-dimensional Euclidean CFT \cite{Anninos:2009yc}. The evidence for the proposal rests in the study of the asymptotic symmetry group \cite{Brown:1986nw} of the rotating Nariai geometry, which is given by a single copy of the Virasoro algebra with a positive central charge. It is found that there is a striking agreement between the various quantities computed for the bulk scalars and those expected from the CFT, upon a suitable identification of the scalar field parameters. In particular, the thermal boundary-to-boundary correlators of the scalar field at $\mathcal{I}^+$ take the form of a two-point function at finite temperature in a two-dimensional CFT. In fact, they imply the presence of both left and right-moving sectors. The right-moving temperature is precisely the Hawking temperature of the cosmological horizon in the $dS_2$ part of the geometry and the left moving temperature is related to the periodicity of the axial coordinate of the black hole. To have complete consistency, we have to also posit the existence of an additional $U(1)$ symmetry whose zero-mode coincides with the zero-mode of the left moving Virasoro.

Some of our discussion bears resemblance to the analogous discussion for the Kerr/CFT correspondence \cite{Guica:2008mu}. On the other hand, we have found clear distinctions between the two. For instance, one can define various vacua for the scalar field in the rotating Nariai geometry. Moreover, one observes cosmological particle production at $\mathcal{I}^+$ as opposed to Schwinger pair-production at the timelike boundary of NHEK geometry. This is reminiscent of the striking difference between scalar fields in de Sitter space, which contain a complex valued parameter worth of vacua, and scalars in anti-de Sitter space which exhibit no such family. Thus, although classically the NHEK and rotating Nariai geometries are related by an analytic continuation in the coordinates, they are significantly different at the quantum level.

\section{Geometry and Conserved Charges}


Our story begins with the four-dimensional Einstein-Hilbert action endowed with a positive cosmological constant
\be
I_g = \frac{1}{16\pi G}\int_{\mathcal{M}} d^4 x \sqrt{-g} \left(R - 2\Lambda \right), \quad \Lambda \equiv + \frac{3}{\ell^2},
\ee
where we set $G = 1$ in what follows. The metric of the rotating black hole in de Sitter space is a two-parameter solution given by
\begin{multline}
ds^2 = - \frac{\Delta_r}{\rho^2} \left( dt - \frac{a}{\Xi} \sin^2\theta d\phi \right)^2  + \frac{\rho^2}{\Delta_r} dr^2 \\ + \frac{\rho^2}{\Delta_\theta}d\theta^2 + \frac{\Delta_\theta}{\rho^2}\sin^2\theta \left( a dt -  \frac{r^2 + a^2}{ \Xi} d\phi \right)^2
\end{multline}
where we have defined the following objects:
\begin{eqnarray}
\Delta_r &=& (r^2+a^2)\left( 1 - \frac{r^2}{\ell^2} \right) -{2 M}{r}, \quad \Xi = 1 + \frac{a^2}{\ell^2},\\
\Delta_\theta &=& 1 + \frac{a^2}{\ell^2}\cos^2\theta, \quad \rho^2 = r^2 + a^2 \cos^2\theta.
\end{eqnarray}
The parameters $a$ and $M$ will be related to the angular momentum and mass of the black hole solution. We will be mostly concerned in the parameter space allowing $\Delta_r$ to contain four (possible repeated) real roots.\footnote{Solutions with two positive real and two complex roots also exist, however such configurations require imaginary $a$ and $M<0$.} We generally write $\Delta_r$ as
\be
\Delta_r = -\frac{1}{\ell^2}(r-r_c)(r-r_+)(r-r_-)(r+r_n), \quad r_n > r_c \ge r_+ \ge r_- >  0
\ee
with the following conditions obeyed:
\be\label{roots}
(r_c + r_+)(r_c +r_-)(r_+ + r_-) = 2M \ell^2, \quad \prod_i r_i = a^2 \ell^2  = - \ell^2 \sum_{i \le j}r_i r_j, \quad r_n = \sum_{i \neq n} r_i.
\ee
We have denoted the locations of the cosmological, outer, inner and negative horizons as $r_c$, $r_+$, $r_-$ and $r_n$ respectively. Note that the negative root $-r_n$ may be physical in the case of rotating black holes given that the singularity is a ring singularity that observers can go through. Furthermore, note that \ref{roots} implies $M > 0$. The de Sitter length is denoted by $\ell$.

\subsection*{Conserved Charges and the First Law}

The conserved charges of the full spacetime have been computed in \cite{Ghezelbash:2004af,Balasubramanian:2001nb} based on the classic construction by Brown and York \cite{Brown:1992br}, and are given by
\be
\mathcal{Q}_{\partial_t} = -\frac{M}{\Xi^2}, \quad \mathcal{Q}_{\partial_\phi} = -\frac{a M}{\Xi^2}\label{cosmcharge}.
\ee
It may seem surprising to find a negative energy however this follows quite naturally from a thermodynamic argument. As we shall see, the most entropic configuration is given by pure de Sitter space which in four-dimensions has vanishing energy. Thus, it is natural that low entropy fluctuations such as the black holes carry less energy than de Sitter space itself.

On the other hand our interest will lie in the thermodynamic properties of the black hole horizon which can be treated as a thermodynamic entity in itself. The energy and angular momentum of the black hole horizon can be defined to be
\be
E \equiv -\mathcal{Q}_{\partial_t} = \frac{M}{\Xi^2}, \quad J \equiv -\mathcal{Q}_{\partial_\phi} = \frac{a M}{\Xi^2}.
\ee
The reason we choose these definitions is that they reduce to the Minkowksi values in the limit $\ell \to \infty$. Additionally, as we shall soon see, they are the ones that give the correct first law of thermodynamics.\footnote{It is an interesting point that the definitions of energy and angular momentum for the black hole are precisely minus those for the full cosmological horizon. It is tempting to view the black hole as inducing an equal and negative energy and angular momentum at the horizon, as an electric charge inside a conducting sphere would \cite{Banks:2005bm}.}

The entropies of the cosmological horizon $S_c$ and black hole $S_{BH}$ are given by
\be
S_c = \frac{\pi(r_c^2 + a^2)}{\Xi}, \quad S_{BH} = \frac{\pi(r_+^2 + a^2)}{\Xi}.
\ee
In the presence of rotation the first law of thermodynamics becomes
\be
d E = T_H d S + \tilde{\Omega}_H d J,
\ee
where $\tilde{\Omega}_H$ and $T_H$ are the angular velocity of the black hole horizon with respect to a non-rotating boundary and Hawking temperature of the black hole. Explicitly, they are given by
\begin{eqnarray}\label{th}
T_H &=& \frac{|\Delta_r' (r_+)|}{4\pi(r^2_+ + a^2)} = \frac{\ell^2 r_+^2 - 3 r^4_+ -a^2 r_+^2 - a^2 \ell^2}{4\pi \ell^2 r_+ (r_+^2 + a^2)}\\ \label{om}
\tilde{\Omega}_H &\equiv& \Omega_H - \Omega_\infty = \frac{a \Xi }{r_+^2 + a^2} - \frac{a}{\ell^2}. 
\end{eqnarray}
Note that the angular velocity as $r \to \infty$ is defined as
\be
\Omega_\infty \equiv -\lim_{r\to\infty}\left( \frac{g_{t\phi}}{g_{\phi\phi}} \right) = \frac{a}{\ell^2}.
\ee

Finally, the temperature and angular momentum of the cosmological horizon, $T_c$ and $\tilde{\Omega}_c$, are given by \ref{th} and \ref{om} with $r_+$ replaced by $r_c$.

\subsection*{The Various Limits of Parameter Space}

There are various regions of interest in the parameter space of the rotating de Sitter black hole. We list them below:
\begin{itemize}
\item The \emph{Extremal Limit}  corresponds to $r_+ \to r_-$ such that the black hole becomes extremal and its temperature vanishes.
\item The \emph{Lukewarm Limit}  corresponds to the black hole and cosmological horizons having the same temperature without necessarily coinciding.
\item The \emph{Rotating Nariai Limit} corresponds to $r_+ \to r_c$ such that the black hole and and cosmological horizons coincide.
\end{itemize}

\begin{figure}[h]\label{configs}
\begin{center}
\includegraphics[angle=0,width=0.6\textwidth]{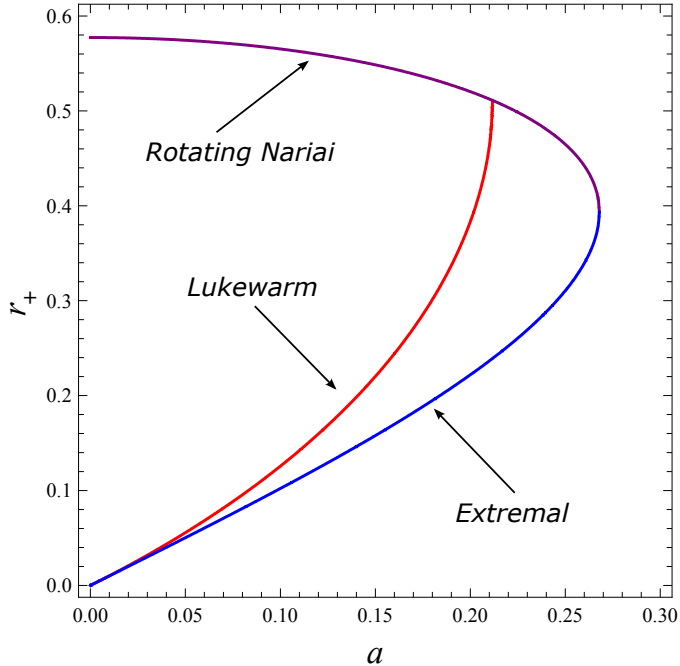}
\end{center}
\caption{The physically allowed configurations for Kerr-de Sitter space. We are using units where $\ell = 1$.}
\end{figure}

It is important to note that in all these limits, one can take a sensible analytic continuation to the Euclidean instanton \cite{Booth:1998gf}. This is in sharp contrast to the generic de Sitter black hole which is out of thermal equilibrium with the cosmological horizon. Particularly, the rotating Nariai instanton is argued to mediate the nucleation of rotating black holes in direct analogy to the original case of the non-rotating Nariai instanton as studied by Ginsparg and Perry \cite{Ginsparg:1982rs} (see also \cite{Bousso:1995cc}).

The lukewarm solution also plays a potentially interesting role as the unique rotating black hole which is in a stable thermal equilibrium with the cosmological horizon at non-zero temperatures. One can obtain an explicit condition for when the black holes are lukewarm, namely
\be
M_{lw} = a\left(1 + \frac{a^2}{\ell^2}\right).\label{lukewarm}
\ee
Having said this, it should also be noted that to attain a system which is in complete thermal equilibrium all thermodynamic chemical potentials must be equal. The angular velocity of the cosmological and black hole horizons for the lukewarm configurations are not equivalent unless we are also at the rotating Nariai limit so in general there will be exchange of particles carrying angular momentum.

\subsection*{Geometry of the Rotating Nariai Limit}

As mentioned above, the rotating Nariai geometry possesses the interesting feature of being in thermal equilibrium with respect to both its temperature and its angular velocity. Here we present the near horizon limit leading to the rotating Nariai geometry.

We will take the Nariai limit $r_+\to r_c$ and the near horizon limit simultaneously. This is the Nariai analog of the near-NHEK limit of extremal black holes considered in \cite{Bredberg:2009pv,Castro:2009jf}. We define the non-extremality parameter
\be
\lambda = \frac{{r_c - r_+ }}{\epsilon r_c}  .
\ee
where $\epsilon$ is a small parameter which we will take to zero. In order to go to the near horizon limit, we must go to a non-rotating frame with respect to the cosmological horizon and rescale the coordinates as follows:
\be
\hat {t} = b \epsilon {t} \ , \quad x = {r - {r_+}\over \epsilon r_c} \ , \quad \hat{\phi} = {\phi} - \Omega_H {t} ,
\ee
where
\be
\Omega_H \equiv {\Xi a\over r_+^2 + a^2}, \quad b \equiv {r_c(r_c-r_-)(3r_c + r_-)\over \ell^2 (a^2 + r_c^2)}.
\ee
Taking $\epsilon \to 0$ with $\lambda$, $t$, $r$, $\phi$ held fixed, we find the rotating Nariai metric \cite{Booth:1998gf}
\be\label{thermalnariai}
ds^2 = \Gamma(\theta)\left(-x(\lambda - x)d\hat{t}^2 + {dx^2\over x(\lambda - x)} + \alpha(\theta)d\theta^2\right) + \gamma(\theta)(d\hat{\phi} + k x d\hat{t})^2 ,
\ee
with $\hat{\phi} \sim \hat{\phi} + 2\pi$, $x \in (0,\lambda)$, and
\be\label{nariaiparam}
\Gamma(\theta) = { \rho_c^2 r_c\over b (a^2 + r_c^2)} , \quad
\alpha(\theta) = {b (a^2 + r_c^2)\over r_c\Delta_\theta} , \quad
\gamma(\theta) = {\Delta_\theta(r_c^2 + a^2)\sin^2\theta\over\rho_c^2 \Xi^2} ,
\ee
\be\notag
k = {2 a r_c^2\Xi\over b (a^2 + r_c^2)^2}, \quad
\rho_c^2 = r_c^2 + a^2 \cos^2\theta .
\ee
At fixed polar angle, one can recognize the above geometry as an $S^1$ fibration over two-dimensional de Sitter space \cite{Anninos:2009jt}. The black hole horizon is located at $x = 0$ and the cosmological horizon is located at $x = \lambda$ and they have the same Hawking temperature
\be \label{trn}
T_{RN} = {\lambda}/{4\pi}.
\ee
Furthermore, both horizons have vanishing angular velocity in the limit $\lambda \to 0$.

\subsubsection*{Global Coordinates}

It will be useful to write down the rotating Nariai geometry in global coordinates. This amounts to writing the $dS_2$ piece in its global form
\be
ds^2 = \Gamma(\theta) \left( -d\tau^2 + \cosh^2\tau d\psi^2 + \alpha(\theta) d\theta^2 \right) + \gamma(\theta)\left(d\phi - k \sinh\tau d\psi  \right)^2
\ee
where $\tau \in (-\infty,\infty)$ and $\psi \sim \psi + 2\pi$ in order to have a single cover of the global $dS_2$. Constant time slices in this spacetime have an $S^1 \times S^2$ topology.

\section{Thermal Phase Structure}

In this section we wish to explore the thermodynamic stability and thermal evolution of the Kerr-de Sitter spacetimes. We begin by discussing stability of the black holes as it arises in the canonical and grand canonical ensembles. We conclude with an evaluation of the thermal evolution based on the total entropy of our system, which we take to be the sum of the cosmological and black hole horizon entropies. Explicit expressions for the objects we compute are presented in appendix \ref{expressions}.

\subsection*{Thermal Stability}

In addition to the first law of thermodynamics, one can study the thermal stability of our system. The measure of thermal stability depends on the ensemble we choose.

\subsubsection*{Canonical Ensemble}

The \emph{canonical ensemble} is defined at a fixed temperature and angular momentum for the black hole. The relevant thermodynamic potential is given by the Helmoltz free energy
\be
\mathcal{F} = E - T_H S_{BH}
\ee
and we must examine the specific heat capacity at fixed angular momentum,
\be
C_J = \left( \frac{\partial E}{\partial T_H} \right)_J = T_H \frac{\partial S_{BH}}{\partial T_H}.
\ee
From the above expression, one notes that both the extremal and rotating Nariai solutions have vanishing specific heat.

In Fig. 2 (a) we exhibit the allowed rotating black hole configurations in the $(r_+,a)$-plane. The black hole horizons with positive specific heat inhabit the region below the dotted line. Notice that for a given angular momentum, there is a phase transition from positive to negative specific heat as one increases $r_+$. This was first observed in \cite{Davies:1989ey} and corresponds to the point where the temperature of the black hole reaches a maximum with respect to $r_+$. The physically allowed parameter space is bounded by the rotating Nariai solutions, and the extremal black hole solutions. The point where the two curves meet is the ultracold point.

We note that most but not all lukewarm configurations have positive specific heat. Extremal and rotating Nariai configurations have vanishing specific heat. Finally, regions where the cosmological horizon has greater (smaller) temperature than the black hole horizon is given by the region above (below) the lukewarm curve.
\begin{figure}[h]\label{phasediag}
\begin{center}
$\begin{array}{c@{\hspace{0.1in}}c}
\includegraphics[angle=0,width=0.5\textwidth]{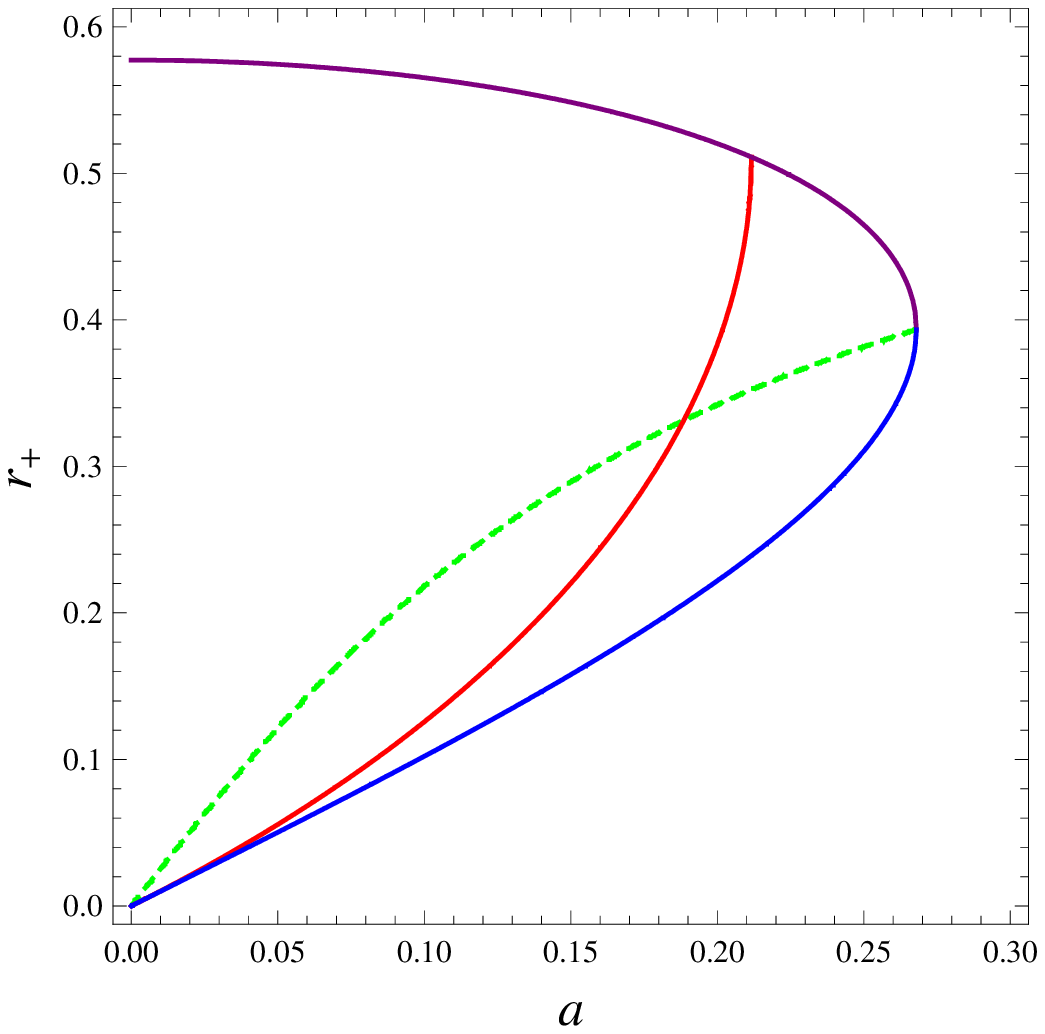} &
 \includegraphics[angle=0,width=0.5\textwidth]{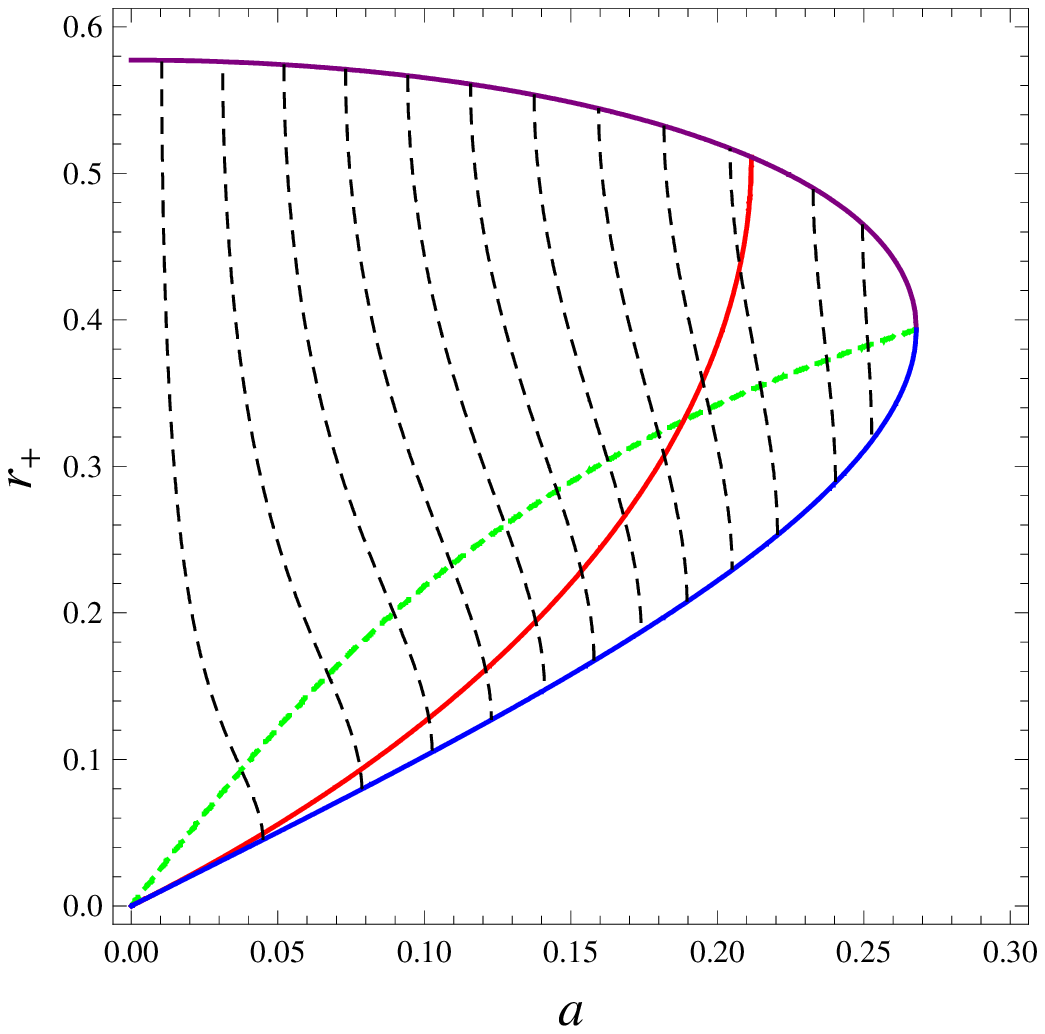} \\
   \text{(a)}&\text{(b)}
   \end{array}$
\end{center}
\caption{(a): Phase space of allowed solutions in the $(r_+,a)$-plane. Above the green (dotted) line, the black hole horizon has negative specific heat. The red (solid) line indicates the lukewarm configurations. (b): Constant $J$ curves in the $(r_+,a)$-plane. We are plotting in units where $\ell = 1$. }
\end{figure}

\begin{figure}[h]\label{phasediag2}
\begin{center}
$\begin{array}{c@{\hspace{0.1in}}c}
\includegraphics[angle=0,width=0.5\textwidth]{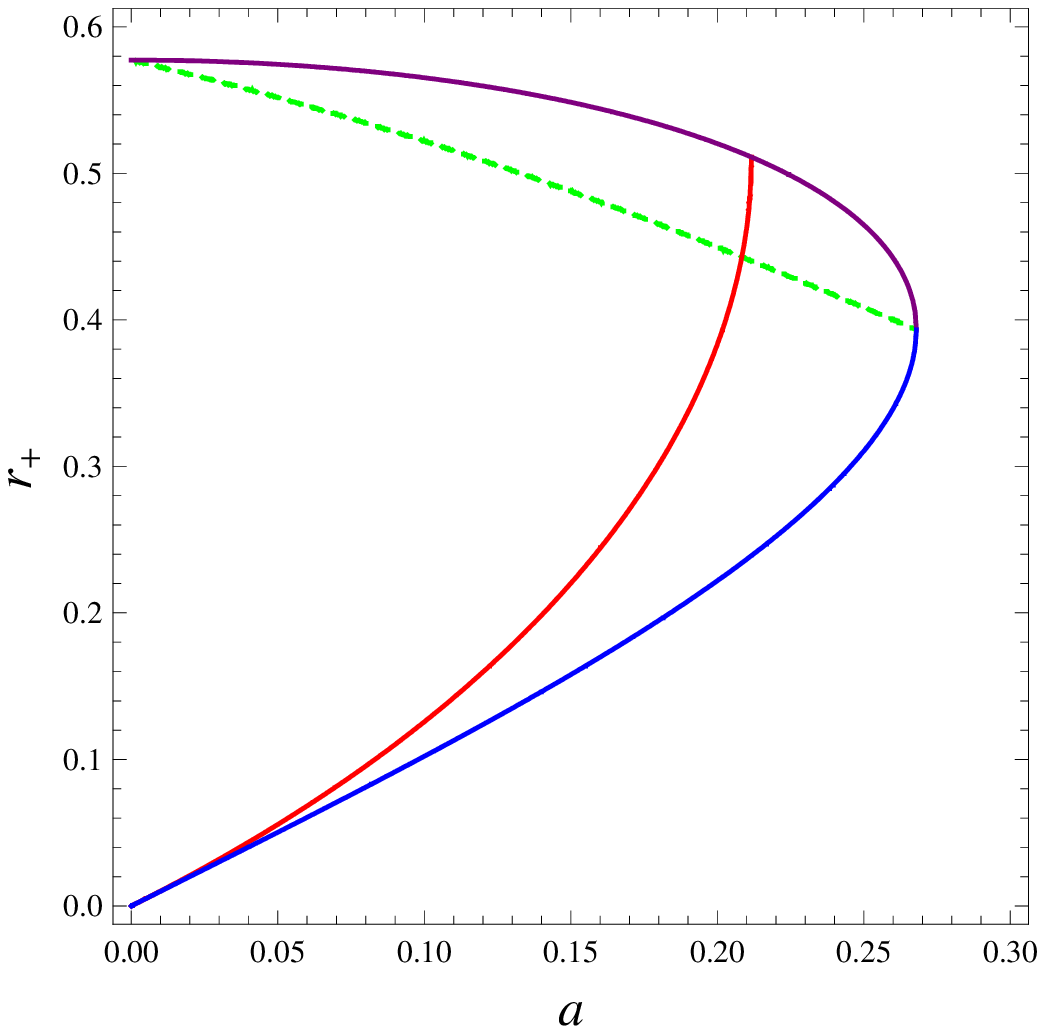} &
 \includegraphics[angle=0,width=0.5\textwidth]{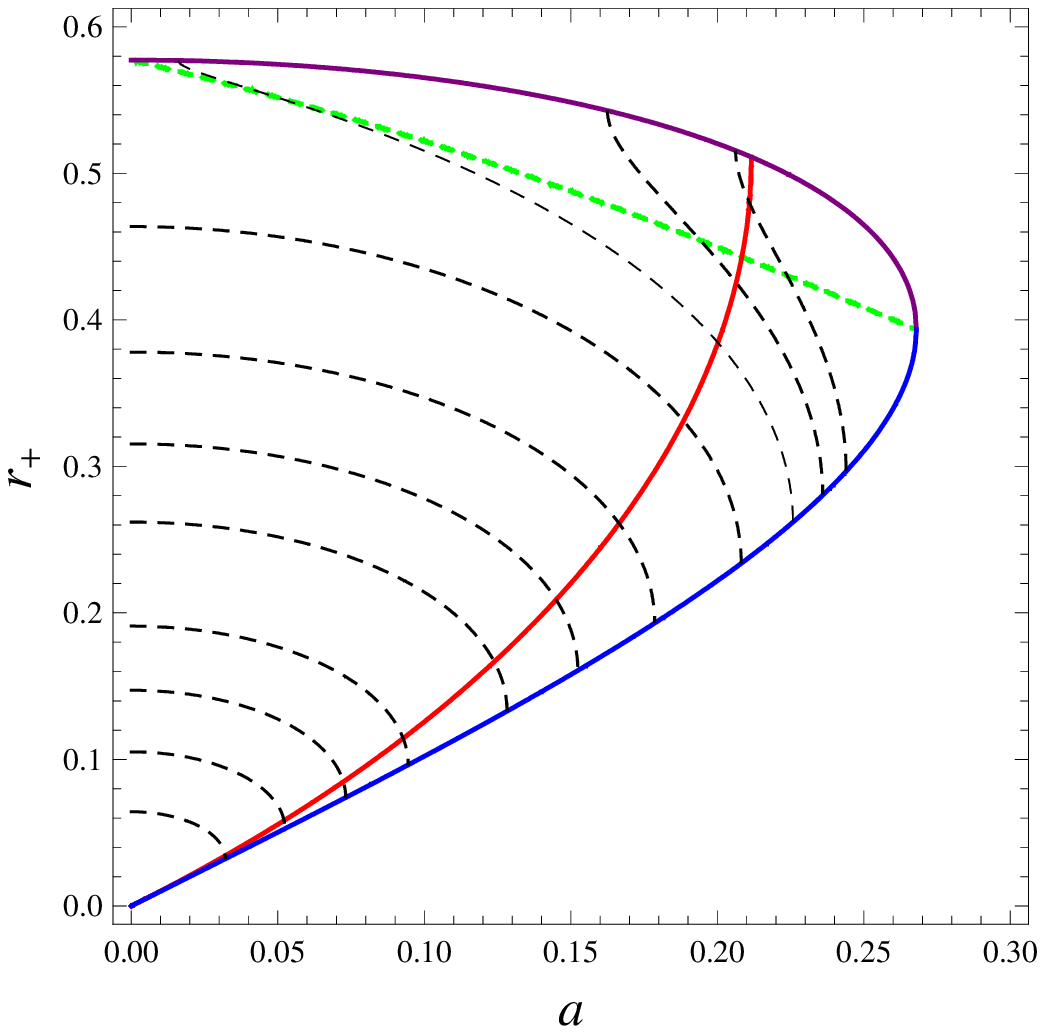} \\
   \text{(a)}&\text{(b)}
   \end{array}$
\end{center}
\caption{(a): Regions of negative $\partial S_{BH}^2/\partial J^2$ below the green (dotted) curve. (b): Constant $E$ curves in the $(r_+,a)$-plane. We are plotting in units where $\ell = 1$. }
\end{figure}

\subsubsection*{Grand Canonical Ensemble}

The \emph{grand canonical ensemble} is defined at a fixed temperature and angular velocity. In this case, the relevant thermodynamic potential is given by the Gibbs free energy,
\be
\mathcal{G} = E - T_H S_{BH} - \tilde{\Omega}_H J
\ee
which is a function of the intrinsic parameters $T_H$ and $\tilde{\Omega}_H$. The stability in the grand canonical ensemble is given by analyzing the full Hessian
\be
H_{ij} = \left( \frac{\partial^2 S_{BH}}{\partial X_i \partial X_j} \right), \quad X_i = E, J.
\ee
A system that is thermally stable will have a total entropy function lying at a maximum, as a function of the extensive parameters. This gives rise to the conditions
\be\label{conditions}
\frac{\partial^2 S_{BH}}{\partial E^2} < 0, \quad \frac{\partial^2 S_{BH}}{\partial J^2} < 0, \quad \frac{\partial^2 S_{BH}}{\partial J^2} \frac{\partial^2 S_{BH}}{\partial E^2} - \left( \frac{\partial^2 S_{BH}}{\partial J \partial E} \right)^2 > 0.
\ee

The first condition is equivalent to the specific heat at fixed angular momentum being positive, the second condition is the analogous statement for fixed energy fluctuations and the third is the requirement that the Hessian have positive determinant. In Fig. 3 (a) we demonstrate the regions of positive and negative $\partial S_{BH}^2/\partial J^2$. It is further found that the Hessian, given explicitly in \ref{hessian} is negative definite for all configuration space, indicating that all black holes are thermally unstable once we allow angular momentum to be exchanged.

\subsection*{Thermal Evolution}

We would like to address the issue of the thermal evolution of the black holes immersed within the cosmological horizon. First note that the cosmological horizon has lower temperature than the black hole horizon when we are in the region above the lukewarm line in Fig. 1. Secondly, the cosmological horizon has an angular velocity that is less or equal to that of the black hole horizon, where equality only holds in the rotating Nariai limit. Thus, most configurations are out of thermal equilibrium and will thermally evolve.

The total entropy of our system is taken to be the sum of the black hole and cosmological entropies,
\be
S_{tot} \equiv S_{BH} + S_c =  \frac{\pi(r_+^2 + a^2)}{\Xi} +  \frac{\pi(r_c^2 + a^2)}{\Xi}.
\ee
Furthermore, the total energy and angular momentum of our spacetime is zero, as it was noticed earlier that the conserved charges of the cosmological horizon are equal and opposite to those of the black hole.

Our system will evolve thermodynamically in the direction that maximizes total entropy for fixed total energy and angular momentum. In Fig. 4 we demonstrate constant $S_{tot}$ contours throughout the configuration space. The system evolves to the pure de Sitter configuration which indeed is the most entropic configuration. Particularly, upon nucleation of the rotating Nariai black hole the two horizons will exchange angular momentum and energy until the black hole spins down and fully evaporates. 
\begin{figure}[h]\label{phasediag3}
\begin{center}
\includegraphics[angle=0,width=0.55\textwidth]{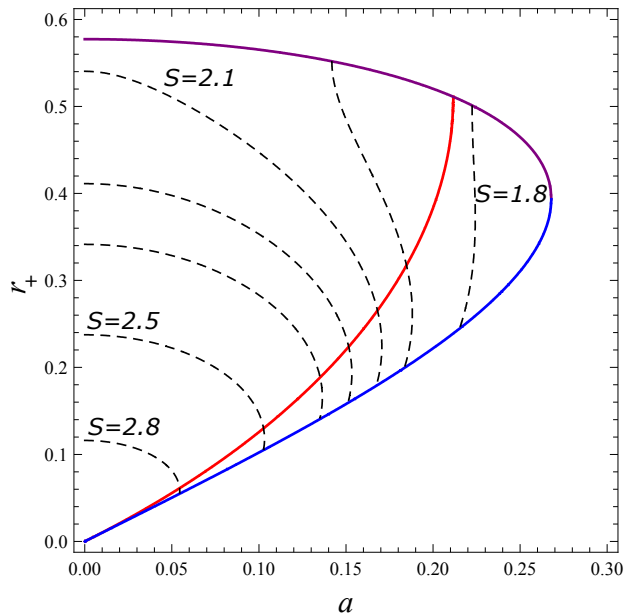}
\end{center}
\caption{Contour plot of constant total entropy curves. The direction of increasing entropy is toward the origin of the configuration space, i.e. pure de Sitter space. We plot in units where $\ell = 1$.}
\end{figure}

In order to determine the direction in which the thermal evolution proceeds throughout our configuration space, we would require knowledge of the emission rates of angular momentum and energy from one horizon to the other (e.g. \cite{Tachizawa:1992ue,Doukas:2009cx}). In appendix \ref{thevol} we present the qualitative possibilities that may appear as we vary the relative rate of emission of energy and angular momentum.

\section{Scalar Waves}

At this point we will turn to the issue of the wave equation for scalar fields \cite{Birrell:1982ix} in the Kerr-de Sitter geometry. We will not be able to obtain explicit solutions in general, the only exception being when we take the rotating Nariai limit. Indeed, the rotating Nariai instanton mediates the semi-classical production of the rotating Nariai geometry \cite{Booth:1998gf} and it is thus a natural configuration to study as the starting point in the thermal evolution of Kerr-de Sitter space. Furthermore, the asymptotic symmetry group of this geometry was recently studied and found to consist of a centrally extended copy of the Virasoro algebra \cite{Anninos:2009yc} suggesting that there may be a holographic interpretation of the spacetime in terms of a two-dimensional conformal field theory.

\subsection*{Scalar Wave Equation}

A simple way to examine the superradiant properties of a rotating black hole are by analyzing a scalar field in the black hole background. The equation of motion for a massless scalar is given by
\be
\nabla^2 \Phi(r,t,\Omega) = 0.
\ee
One can impose an ansatz for which the variables separate and obtain an angular and radial equation. If we choose $\Phi(r,t,\Omega) = R(r)Y_{l m}(\theta)e^{-i \omega t + i m \phi}$, with $m \in \mathbb{Z}$, we find two decoupled equations for the angular and radial parts.

The angular equation is given by the spheroidal harmonic equation
\be
\left( \frac{1}{\sin\theta}\frac{d}{d\theta} \Delta_\theta \sin\theta \frac{d}{d\theta} -  \frac{\left(m \Xi - a\omega \sin^2\theta\right)^2}{\Delta_\theta \sin^2\theta} +  j_{lm} \right) Y_{lm}(\theta) = 0.
\ee
It is not hard to prove that the separation constants $j_{lm}$ are discrete for a given $\omega$, however their values are only known numerically. Our normalization for the $Y_{lm}$ is given in appendix \ref{twopoint}. The radial equation is given by
\be
\left( \frac{d}{dr} \Delta_r \frac{d}{dr} + \frac{1}{\Delta_r}\left( a  \Xi  m - \omega (a^2 + r^2) \right)^2  -  j_{lm} \right) R(r)= 0.
\ee

The above equations are generally not analytically solvable. They are, however, regions where an approximate analytic grasp is possible. For instance, when the black hole is near-extremal, i.e. $r_+ \to r_-$, we can solve the wave-equation in the near horizon region. More precisely, if we define the parameter
\be
x = \frac{r-r_+}{ r_+}
\ee
and a unitless Hawking temperature $\tau_H \propto (r_+ - r_-)$ then one can match the $x \gg \tau_H$ and $x \ll 1$ solutions in the region $ \tau_H \ll x \ll 1$ and compute the reflection coefficient. This situation has been discussed for the near-extremal Kerr black hole \cite{Bredberg:2009pv}, and we will not pursue it here.

\subsection*{Rotating Nariai Limit}

We will instead focus on the rotating Nariai limit. In this limit, the black hole and cosmological horizons coincide and we must take a near horizon limit to reveal the underlying geometry where observers reside. The small parameters relevant to this limit are the near extremality parameter $\lambda$ and the near horizon variable $x$ defined by
\be
x = \frac{(r - r_+)}{\epsilon r_c}, \quad \lambda = \frac{(r_c - r_+)}{\epsilon r_c}.
\ee

In the limit $\epsilon \to 0$ with $\lambda$ fixed, we find that our radial equation tends to
\be\label{DiffeqRotNar}
\left( x(x-\lambda)\frac{d^2}{dx^2} +  (2x-\lambda)\frac{d}{dx} +  \frac{\left( \tilde{\omega} + m k x  \right)^2}{x(x-\lambda)} + \tilde{j}_{lm}  \right) R(x) = 0,
\ee
where we have defined
\be
k \equiv \frac{a r_c \Xi }{(r_c^2+a^2)}\frac{2\ell^2}{(3r_c+r_-)(r_c-r_-)}, \quad \tilde{j}_{lm} \equiv \frac{j_{lm} \ell^2}{(3r_c + r_-)(r_c - r_-)}.
\ee
We note that in order to obtain the above equation, we have to restrict ourselves to frequencies satisfying the `near superradiant bound'
\be
\omega = m \Omega_H + \epsilon \tilde{\omega}  \frac{{r_c(r_c - r_-)(3r_c + r_-)}}{\ell^2(a^2 + r_c^2)}, \quad \Omega_H \equiv \frac{a\Xi}{r_+^2+a^2}
\ee
and the sign of $\tilde{\omega}$ determines if we are above or below the bound $\omega = m\Omega_H$. In this limit the black hole horizon resides at $x=0$ and the cosmological horizon resides at $x = \lambda$.

It is interesting to note that $R(x)$ obeys the equation of motion for the radial part of a charged massive scalar field in two-dimensional de Sitter space (with de Sitter radius $\ell_2$), in the presence of an $\bold{E}$-field \cite{Kim:2008xv}. Explicitly, the $\bold{E}$-field and mass $m_2^2$ are related to the four-dimensional quantities by:
\be
q \bold{E} = m k, \quad m_2^2 \ell_2^2 = \tilde{j}_{lm}.
\ee
Furthermore, we identify the radial equation as the equation satisfied by the radial part of a massive scalar field in the rotating Nariai geometry \ref{thermalnariai}.  The frequency $\tilde{\omega}$ becomes the frequency of the modes in the $\hat{t}$-coordinate, thus giving a time dependence of the form $e^{-i\tilde{\omega}\hat{t}}$.

The solution to this equation is given by a linear combination of hypergeometric functions. An explicit solution in the region $0 < x < \lambda$ is given by
\be
R(x) = (\lambda - x)^{{-i n_R }}\left[ c_1 \times x^{i{\tilde{\omega}}/{\lambda}} \; R_1(x) + c_2 \times x^{-i{\tilde{\omega}}/{\lambda}}  \; R_2(x) \right]
\ee
where
\begin{eqnarray}
R_1 (x) &=&    {_2}F_1\left(h_- - i m k,h_+ - i m k,1+{2i\tilde{\omega}}/{\lambda}, {x}/{\lambda}\right), \\
R_2 (x) &=&    {_2}F_1\left(h_- - i (n_R + \tilde{\omega}/{\lambda}),h_+ - i(n_R + \tilde{\omega}/{\lambda}),1-{2i\tilde{\omega}}/{\lambda}, {x}/{\lambda}\right).
\end{eqnarray}
We have defined the following convenient quantities
\begin{eqnarray}
h_\pm &\equiv& \frac{1}{2} \pm   i \frac{\mu}{2}  , \quad  \mu = {\sqrt{4 \tilde{j}_{lm} + 4 m^2 k^2 - 1}},\\
n_R &\equiv& m k+\tilde{\omega}/\lambda,
\end{eqnarray}
and we assume for later convenience that $i\mu = h_+ - h_-$ (when real) is non-integral. The expressions $h_{\pm}$ are complex for large angular momentum along the two-sphere. This property will be related to the oscillatory behavior of modes at $\mathcal{I}^{\pm}$. Also, it is worth noting that $h_{\pm}$ have an implicit dependence on $\tilde{\omega}$ through the $\tilde{j}_{l m}$.

We now proceed to explore the above solution in the various interesting regions of the geometry: the horizons and the boundary of the rotating Nariai geometry.

\subsubsection*{Behavior Near the Cosmological Horizon}

Demanding that our wavepackets have ingoing group velocity at the black hole horizon leads us to impose $c_1 = 0$. Having done so, we can expand our solution near the cosmological horizon\footnote{In what follows we perform each expansion using hypergeometric function identities found in Abramowitz \& Stegun \cite{Abramowitz:1965}.} and find a linear combination of ingoing and outgoing waves\footnote{The subscripts $out$, $in$ refer to the direction in which positive energy flux is traveling, i.e. the $out$ modes have positive flux escaping the cosmological horizon. These need not coincide with the direction of the group velocity. We define flux in \ref{flux}.}
\be
R(x) = \left[ A_{out} (\lambda - x)^{-i n_R} + A_{in} (\lambda - x)^{i n_R }   \right] \label{cosmodes}
\ee
where the ingoing and outgoing coefficients are given by
\begin{eqnarray}
A_{out} &=& \frac{\Gamma[1-{2i\tilde{\omega}}/{\lambda}]\Gamma[2i n_R]\lambda^{-i\tilde{\omega}/\lambda}}{\Gamma[h_+ + i m k]\Gamma[h_- + i m k]},\\
A_{in} &=& \frac{\Gamma[1-{2i\tilde{\omega}}/{\lambda}]\Gamma[-2i n_R]\lambda^{-i(2  n_R+\tilde{\omega}/\lambda)}}{\Gamma[h_+ - i (n_R +\tilde{\omega}/\lambda)]\Gamma[h_- - i (n_R +\tilde{\omega}/\lambda)]}.
\end{eqnarray}
It is important to note that in the above, we are implicitly considering the case where $\Re n_R > 0$. In the case where $\Re n_R < 0$, one must switch the labels $A_{in}$ and $A_{out}$.

\subsubsection*{Behavior at Late/Early Times}
We now want to consider the behavior of scalar waves in the rotating Nariai geometry as they approach the asymptotic past or future, i.e. $ \lambda < x < \infty$. In order to do so we introduce the following two linearly independent solutions to \ref{DiffeqRotNar}:
\begin{eqnarray}
R_-(x) = x^{ - h_- + i n_R  }\left( x - \lambda \right)^{-i n_R} {_2}F_1 \left( h_- - i( n_R +\tilde{\omega}/\lambda), h_- - i m k, 2 h_-, \lambda/x \right),\\
R_+(x)  = x^{ - h_+ + i n_R  }\left( x - \lambda \right)^{-i n_R} {_2}F_1 \left( h_+ - i m k, h_+ - i (n_R +\tilde{\omega}/\lambda), 2 h_+, \lambda/x \right).
\end{eqnarray}
Note that in the large $x$ limit, the above solutions behave as:
\be
R_{\pm}(x) \sim x^{ - h_\pm}.
\ee
Given that $x$ becomes timelike in the region $x>\lambda$, $R_+(x)$ corresponds to positive frequency modes and $R_- (x)$ corresponds to negative frequency modes for real $\mu$ at $\mathcal{I}^+$, where increasing $x$ corresponds to the direction of increasing time. Note that $\hat{t}$ is now a spacelike variable whose range is $[-\infty, \infty]$ in the region $x>\lambda$. Forward spatial motion in this patch is defined along the direction of decreasing $\hat{t}$.

As in the previous subsection, we can expand both $R_-$ and $R_+$ near $x = \lambda$ using hypergeometric function identities. We find
\be\label{rminus}
R_-(x) = x^{ - h_- + i n_R  }\left( B_{out}^- \left( x - \lambda \right)^{-i n_R} + B_{in}^- (x - \lambda)^{i n_R}x^{-2i n_R} \right),
\ee
where the coefficients are found to be
\begin{eqnarray}
B_{out}^- &=& \frac{\Gamma[2 h_-]\Gamma[2 i n_R]}{\Gamma[h_- + i m k]\Gamma[h_- + i( n_R + \tilde{\omega}/\lambda)]},\\
B_{in}^- &=& \frac{\Gamma[2 h_-]\Gamma[-2 i n_R]}{\Gamma[h_- - i m k]\Gamma[h_- - i (n_R + \tilde{\omega}/\lambda)]}.
\end{eqnarray}
Similarly, the $R_+$ modes near $x = \lambda$ are given by
\be\label{rplus}
R_+(x) = x^{ - h_+ + i n_R  }\left( B_{out}^+ \left( x - \lambda \right)^{-i n_R} + B_{in}^+ (x - \lambda)^{i n_R}x^{-2i n_R}\right),
\ee
where the coefficients are found to be
\begin{eqnarray}
B_{out}^+ &=& \frac{\Gamma[2 h_+]\Gamma[2 i n_R]}{\Gamma[h_+ + i m k]\Gamma[h_+ + i (n_R + \tilde{\omega}/\lambda)]},\\
B_{in}^+ &=& \frac{\Gamma[2 h_+]\Gamma[-2 i n_R]}{\Gamma[h_+ - i m k]\Gamma[h_+ - i (n_R + \tilde{\omega}/\lambda)]}.
\end{eqnarray}

\subsubsection*{Matching the Flux}
At this point, we would like to match the outgoing flux across the future cosmological horizon. We begin by defining the general solution for $x>\lambda$:
\be\label{rtot}
R_{tot}(x)=\alpha R_- (x) + \beta R_+ (x).
\ee
Matching the outgoing flux across the future cosmological horizon $x = \lambda$, amounts to matching the {\em out} coefficient in \ref{cosmodes} with the {\em out} coefficients in \ref{rtot}. More precisely, we would like to solve the following set of equations
\begin{eqnarray}
\alpha B_{out}^- + \beta B_{out}^+ &=& A_{out},\label{fluxCons}\\
\alpha B_{in}^- + \beta B_{in}^+ &=& 0\label{bndryflux}
\end{eqnarray}
where \ref{bndryflux} implies no localization of flux on the future cosmological horizon. Explicit expressions for $\alpha$ and $\beta$ are provided in appendix \ref{alphabeta}. For future reference, however, we would like to note the ratio of these coefficients:
\be\label{ab}
\frac{\alpha}{\beta} =  \lambda^{-i \mu} \frac{\Gamma[i\mu]\Gamma[h_- - i m k]\Gamma[h_- - i( n_R+ \tilde{\omega}/\lambda)]}{\Gamma[ - i \mu]\Gamma[h_+ - i m k]\Gamma[h_+ - i(n_R+ \tilde{\omega}/\lambda)]}.
\ee

\section{Dissipation and Correlation}

Given the explicit form of the solution, it is natural to study two objects. The first is related to the dissipative properties of the thermal background, which are encoded in the quasinormal modes. The second object we will study is the thermal boundary-to-boundary correlator at $\mathcal{I}^\pm$. The motivation for studying such correlators stems from the possibility that there exists a holographic theory living at the $\mathcal{I}^{\pm}$ boundary \cite{Anninos:2009yc}.

\subsubsection*{Quasinormal Modes of Rotating Nariai}

As was noted, having found the solution in the rotating Nariai geometry we can obtain the quasinormal modes due to scalar fluctuations. These are obtained imposing that the scalar wave has purely ingoing flux at the black hole horizon and purely outgoing flux at the cosmological horizon. As we mentioned, they encode the dissipative behavior of the thermal background under scalar perturbations.

For $\Re n_R > 0$ and $\Re \tilde{\omega} > 0$, this amounts to restricting the values of the angular momentum to the following discrete set
\begin{eqnarray}
{\tilde{\omega}} &=&  - {i \lambda}  (n + h_\pm)/2 - {\lambda m k}/{2}, \quad n = 0,1,2,3,\ldots \label{qnm1} \\
&=& - {2 \pi i T_{RN}}  \left(n + h_\pm \right) - {\lambda m k}/{2}, \quad n = 0,1,2,3,\ldots
\end{eqnarray}
since they would lead to a vanishing $A_{in}$ coefficient in \ref{cosmodes}. The imaginary part of the quasinormal modes is clearly related to the temperature of the horizons \ref{trn}.

For $\Re n_R < 0$ and $\Re \tilde{\omega} > 0$, we find that the quasinormal modes become
\be
|m| k  = - i\left( n + h_\pm \right), \quad n = 0,1,2,3,\ldots \label{qnm2}
\ee
It is important to note that the set of modes \ref{qnm2} imposes a condition on the frequencies $\tilde{\omega}$ through the implicit dependence of $h_{\pm}$ on $\tilde{\omega}$. When $m =0$, the quasinormal modes \ref{qnm1} reduce to those of the non-rotating Nariai geometry \cite{Cardoso:2003sw}.


\subsubsection*{Two-Point Functions: Thermal Background}

According the the notion that there is a holographic dual living at $\mathcal{I}^+$ \cite{Anninos:2009yc}, it is natural to obtain the retarded thermal boundary-to-boundary correlators \cite{Klebanov:1999tb,Son:2002sd} at $\mathcal{I}^+$ of the near horizon region \ref{thermalnariai}.

By imposing the boundary condition that our excitations are purely incoming at the horizon we fix the behavior of the scalar field at $\mathcal{I}^+$. The thermal boundary-to-boundary two-point function is defined by taking variational derivatives of the action with respect to the boundary value $\Phi_0$ of the scalar field
\be
G^{th}_R(q,q') \equiv  \frac{\delta}{\delta \Phi_0 (q)}\frac{\delta I_{matter}}{\delta \Phi_0(q')}, \quad q, q' \in \{ \tilde{\omega}, m, l \}.
\ee
and the matter action for the scalar field is given by the expression\footnote{Further details of the derivation are given in appendix \ref{twopoint}.}
\be
I_{matter} =  \frac{1}{2}\int_{\mathcal M} d^4 x \sqrt{-g} \partial_\mu \Phi \partial^\mu \Phi.
\ee

As we observed earlier, the late time behavior of the scalar field (in momentum space) about the thermal background is given by
\be
\Phi \sim \alpha x^{- h_+} + \beta x^{-h_-}.
\ee
Modes with either of the falloffs $h_\pm$ are normalizable with respect to the Klein-Gordon inner product given by,
\be
(\Phi_1, \Phi_2) = -i \int_{\Sigma} d^3x \sqrt{h} n^\mu \left( \Phi_1 \overleftrightarrow{\partial_\mu} \Phi_2^* \right)\label{klein}
\ee
where $\Sigma$ is a constant time slice with unit normal vector $n^\mu$ and $h_{ij}$ is the induced metric on $\Sigma$. Thus, we have the freedom to choose whether we are `sourcing' an operator with conformal weight $h_+$ or $h_-$, i.e. whether we take variational derivatives of the action with respect to $\beta$ or $\alpha$ as the boundary value. The two different choices lead to the following two Green's functions
\be
G^{th}_R \sim \frac{\alpha}{\beta} \quad \text{or} \quad G^{th}_R \sim \frac{\beta}{\alpha}\label{green}
\ee
for a conformal weight $h_+$ or $h_-$ respectively. The ratio $\alpha/\beta$ was given in \ref{ab}.

\section{Superradiance/Cosmological Particle Production}

Rotating black holes are known to superradiate. Classically, this means that an incoming wave toward the black hole horizon will be reflected back from the gravitational potential with a reflection coefficient larger than unity. Quantum mechanically it gives rise to spontaneous emission of radiation from the black hole horizon carrying angular momentum.

We can get a basic idea of the process by considering the heat transfer $T_H dS$ of a black hole upon the scattering of a quantum with energy $\omega$ and angular momentum $m > 0$. The first law of thermodynamics tells us that
\be
T_H \delta S_{BH} \approx \delta E_{BH} \left(  1 -   \frac{m}{\omega}  \Omega_{H} \right)
\ee
giving us a simple condition $\omega <  m \Omega_{H}$ for the extraction of energy from the black hole.

It has been further noted \cite{Tachizawa:1992ue,Khanal:1983vb} that the presence of a cosmological horizon introduces another condition for superradiance. Namely, given the conserved charges of the cosmological horizon \ref{cosmcharge}, one obtains a first law for the cosmological horizon. The crucial difference with the black hole horizon is that there is a relative sign between the charges which leads to the following condition for the onset of superradiance
\be
T_c \delta S_{c} \approx \delta E_{c} \left(  - 1  +  \frac{m}{\omega} \Omega_{c} \right)
\ee
which leads to the relation $\omega >  m \Omega_{c}$. Clearly, this condition is only relevant when the incoming wave is sent from a region near the cosmological horizon.  

\subsection*{Superadiance in the Rotating Nariai Limit}

We can effectively analyze superradiance in the rotating Nariai limit using our analysis of the scalar wave. We choose boundary conditions such that we have an incoming wave originating near the cosmological horizon which is purely ingoing at the black hole horizon. The flux is given by
\be\label{flux}
f = \frac{1}{2i} \left[ x(\lambda - x)R^* \frac{d}{d x} R -  x(\lambda - x) R \frac{d}{d x} R^* \right].
\ee

The absorption cross-section of the black hole is given by the ratio of the absorbed flux at the black hole horizon to incoming flux from the cosmological horizon and is found to be
\be
\sigma_{abs} = \frac{f_{abs}}{f_{in}} =  \frac{2\sinh(2\pi\tilde{\omega}/\lambda)\sinh(2\pi n_R)}{\cosh(2\pi (n_R + \tilde{\omega}/\lambda)) + \cosh(\pi \mu ) }.
\ee
Thus, when $- m k \lambda < \tilde{\omega} < 0$ the absorption cross-section becomes negative and our system exhibits superradiance. This agrees with our original definition of $\tilde{\omega}$, since it is precisely the deviation away from the superradiant bound. Thus, we can recover the condition on the original frequency $\omega$:
\be
m \Omega_c < \omega < m \Omega_{H}.
\ee
Notice that as $\lambda \to 0$, with $\tilde{\omega}$ and $m$ fixed, the absorption cross-section tends to unity, implying that all incoming radiation is absorbed by the black hole horizon and thus superradiance is absent.

\subsection*{Particle production in the Rotating Nariai Limit}

Given that we are in a cosmological spacetime, we must also investigate the production of particles at late times starting from a given initial vacuum state \cite{Chernikov:1968zm,Mottola:1984ar}. The appropriate metric to address this question is the global metric given by
\be
ds^2 = \Gamma(\theta) \left( -d\tau^2 + \cosh^2\tau d\psi^2 + \alpha(\theta) d\theta^2 \right) + \gamma(\theta)\left(d\phi - k \sinh\tau d\psi  \right)^2,
\ee
where $\tau \in [-\infty,\infty]$ and $\psi \sim \psi + 2\pi$. Notice that this metric contains no horizons, and no single observer can fully access it.

Choosing an ansatz of the form $\Phi(\tau,\psi,\Omega) = T(\tau)Y_{lm}(\theta)e^{i(q\psi + m\phi )}$ with $m$ and $q$ being integers, we find that $T(\tau)$ obeys
\be
\left( \frac{d^2}{d\tau^2} + \tanh\tau \frac{d}{d\tau} +   {\text{sech}^2\tau} \left(q + m k \sinh\tau \right)^2 + \tilde{j}_{lm}  \right) T(\tau) = 0.
\ee
If we perform the coordinate transformations $t = \sinh\tau$ and subsequently $z = t - i$, the solution is found to be
\be
T(z) =  \left( z + 2i \right)^{  (- i \tilde{n} + 2q) /2}    \left( c_1  \times z^{ i \tilde{n} /2 } \; T_1(z)  +  c_2 \times  z^{- i \tilde{n}/2 } \; T_2(z) \right)\label{globsca}
\ee
where the expressions for $T_1(z)$ and $T_2(z)$ are:
\begin{eqnarray}
T_1(z) &=& {_2}F_1 \left( h_- + q, h_+ + q, 1 + i \tilde{n}, i z / 2 \right),\\
T_2(z) &=& {_2}F_1 \left( h_- - i m k, h_+ - i m k, 1 - i \tilde{n}, i z/2 \right)
\end{eqnarray}
and we have defined $\tilde{n} \equiv m k - i q$.

We can obtain the form of the solution for early times, $t \to -\infty$, by using hypergeometric identities. We choose $c_1$ and $c_2$ such that
\be
T_{in}(t) = \frac{(-t)^{-h_-}}{\sqrt{\mu}} \label{inmodes},
\ee
where we have normalized with respect to the Klein-Gordon inner product \ref{klein}. Notice that when $h_-$ becomes complex and thus acquires a negative imaginary part, and we obtain an incoming particle state with positive frequency at $\mathcal{I}^-$. Thus we can expand the $\text{in}$-modes as a sum of creation an annihilation operators of the $|\text{in}\rangle$ vacuum:
\be
\Phi_{in}(t,\psi,\Omega) = \sum_{n \in \{ m,l,q \}} \left( \Phi^{(n)}_{in}(t,\psi,\Omega) a_{in,n} + \Phi^{(n)*}_{in}(t,\psi,\Omega) a^\dag_{in,n} \right)
\ee
with
\be
\Phi^{(n)}_{in}(t,\psi,\Omega) = T_{in}(t) Y_{lm}(\theta) e^{i ( m \phi+ q \psi )}.
\ee
normalized by the Klein-Gordon inner product \ref{klein}. The creation and annihilation operators obey the usual commutation relations with the following normalization
\be
[a_{in,n},a^\dag_{in,m}] = \delta_{nm}, \quad [a_{in,n},a_{in,m}] = 0, \quad [a^\dag_{in,n},a^\dag_{in,m}] = 0.
\ee
Furthermore, the $a_{in,n}$ operators annihilate the $| \text{in} \rangle$ vacuum, i.e. $a_{in,n} | \text{in} \rangle = 0$.

The choice of $c_1$ and $c_2$ giving rise to the purely incoming particle state at past infinity \ref{inmodes} gives rise to the following behavior at future infinity $\mathcal{I}^+$
\be
\lim_{t \to +\infty} T_{in}(t) = b_+ \left( \frac{t^{-h_+}}{\sqrt{\mu}} \right) + b_-  \left( \frac{t^{-h_-}}{\sqrt{\mu}} \right).
\ee
Thus, if we define $| \text{out} \rangle$ as the vacuum state with no outgoing particles on future infinity, which is annihilated by modes of the form
\be
T_{out} (t) = \frac{t^{-h_+}}{\sqrt{\mu}}
\ee
we find the following Bogoliubov transformation
\be
a_{out,n} = b_+ a_{in,n} + b_-^* a^\dag_{in,n}.
\ee
Thus, cosmological particle production due to the fact that $b_-$ is non-vanishing. In other words,
\be
| \text{in} \rangle \neq | \text{out} \rangle.
\ee
The expectation value of the number of $\text{out}$-particles produced in the $| \text{in} \rangle$ vacuum is given by
\begin{eqnarray}
\langle \text{in} | a^\dag_{out,n} a_{out,n} | \text{in} \rangle &=& |b_-|^2 \\
&=& \cosh^2 \left( \pi  m k  \right) \text{csch}^2 \left( \frac{\pi \mu}{2} \right)
\end{eqnarray}
and one can check explicitly that the relation
\be
|b_+|^2 - |b_-|^2 = 1
\ee
is satisfied.\footnote{We work in the Riemann sheet with $-\pi \le \text{Arg} z < \pi$ such that $e^{-i\pi} = -1$.} As a consistency check, one may observe that for $m \to 0$, the result tends to that of cosmological particle production in $dS_2$.

Since the form of the wave equation is qualitatively similar for the rotating Nariai geometry in any number of dimensions, we don't expect this result to be sensitive to the dimensionality of our spacetime. This is in contrast to the regular de Sitter geometry which only exhibits particle production in \emph{even} dimensions \cite{Bousso:2001mw}.

\subsubsection*{Euclidean Modes - A Proposal}

We would like to explore one last vacuum in the global coordinates which we will call the Euclidean vacuum. In regular de Sitter space, it is well known that there exists a family of de Sitter invariant vacua known as the $\alpha$-vacua, which are parameterized by the complex parameter $\alpha$ \cite{Chernikov:1968zm,Mottola:1984ar,Allen:1985ux}. The $\alpha$-vacuum modes are given by a Bogoliubov transformation of the in-modes. The corresponding Green's function in the $\alpha$-vacuum has a singularity both along null separations as well as separations on antipodal points of the sphere, which are separated by a horizon for any given observer.

There is a particular value of $\alpha$ for which the modes become analytic in the lower hemisphere of the Euclidean de Sitter geometry, which is of course $S^{d+1}$, and the vacuum defined becomes the CPT invariant Euclidean vacuum $|E\rangle$. This is the unique $\alpha$-vacuum that reduces to the Minkowski vacuum at short distances. The boundary-to-boundary two-point function in the Euclidean vacuum behaves as that of a $d$-dimensional Euclidean conformal field theory at zero temperature \cite{Strominger:2001pn}.

In a similar fashion, we would like to define the positive frequency Euclidean modes $\Phi^E_n$ in the global rotating Nariai geometry as those which are analytic in the lower hemisphere of the $S^2$ arising from the Euclideanization of the $dS_2$ part of the geometry. A motivation for this definition is the physical relation between the rotating Nariai geometry and $dS_2$ in the presence of an \textbf{E}-field, as was previously noted. Furthermore, they reduce to the Euclidean modes (without the $Y_{lm}(\theta)$) for a massive scalar in $dS_2$ in the limit $m \to 0$.

Let us analytically continue $\tau$ to $i \vartheta$ such that our $z$-variable in \ref{globsca} becomes
\be
z \to z_E = i (  \sin\vartheta - 1 ).
\ee
The upper and lower hemispheres of the $S^2$ are parameterized by $\vartheta \in [0, \pi/2]$ and $\vartheta \in [-\pi/2,0)$ respectively. The argument of the solution in global coordinates then becomes $iz/2 \to (-\sin\vartheta + 1)/2$ which in turn becomes unity at the pole of the lower hemisphere. Thus, we order for \ref{globsca} to be analytic in the lower hemisphere, we have to take a linear combination given by
\be
c_1 = - 2^{- i \tilde{n}} \; e^{- \pi \tilde{n}/2} \times \frac{\Gamma[h_+ + i m k]\Gamma[h_- + i m k]\Gamma[1 - i \tilde{n} ]}{\Gamma[h_+ - q]\Gamma[h_- - q]\Gamma[1+ i \tilde{n}]}  \; c_2  .
\ee
Thus, we can obtain the Wightman function in the Euclidean vacuum as usual
\be
G^{Euc}_W(x,x') =  \langle E | \Phi^E (x)\Phi^E(x') | E \rangle = \sum_{n \in \{l, m, q \}} \Phi^E_n(x) \Phi^{E*}_n(x').
\ee
We hope to study the Euclidean modes and more generally the possibility of $\alpha$-vacua in the rotating Nariai geometry in a future work.

\section{The rotating Nariai/CFT Correspondence}\label{nariaicft}

Having discussed various properties of the Kerr-de Sitter geometry and in particular the rotating Nariai limit, we would like to make contact with the proposal that quantum gravity in a rotating Nariai background is holographically dual to a two-dimensional Euclidean conformal field theory living at $\mathcal{I}^+$.

\subsection*{Asymptotic Symmetries}

In \cite{Anninos:2009yc} it was shown that upon defining suitable boundary conditions, the asymptotic symmetry group of the extremal rotating Nariai geometry was given by a single centrally extended Virasoro algebra
\be
[L_m,L_n] = (m-n)L_{m+n} + m(m^2-1)\frac{c_L}{12}\delta_{m,-n}
\ee
with positive central charge given by,
\be
c_L = \frac{12 r_c^2\sqrt{(1-3r_c^2/\ell^2)(1+r_c^2/\ell^2)}}{-1+6r_c^2/\ell^2+3r_c^4/\ell^4},
\ee
and the $L_n$ are the generators of the algebra. Assuming unitarity and modular invariance, the Cardy formula was then used to compute the cosmological entropy of the extremal Nariai geometry
\be
S_{c} = \frac{\pi^2}{3} T_L c_L, \quad T_L = \frac{1}{2\pi k}\label{cardy}
\ee
where the reciprocal of the left-moving temperature was precisely the chemical potential conjugate to the angular momentum,
\be
dS_c = \frac{1}{T_L} d \mathcal{Q}_{\partial_\phi}.
\ee

\subsection*{Finite Temperature Two-point Function}

One of the most generic features of a two-dimensional conformal field theory is given by the structure of its two-point functions at finite temperature. More precisely, the thermal two-point function in Euclidean momentum-space is given by \cite{Cardy:1984rp}
\begin{multline}
G_E(\omega_L,\omega_R) \sim T_L^{2h_L-2}T_R^{2h_R-2} \Gamma\left[h_L +  \frac{\omega_L}{2\pi T_L}\right] \Gamma\left[h_L -  \frac{\omega_L}{2\pi T_L}\right]  \times \\    \Gamma\left[h_R +  \frac{\omega_R - i q_R \Omega_R}{2\pi T_R}\right] \Gamma\left[h_R -  \frac{\omega_R - i q_R \Omega_R}{2\pi T_R}\right]
\end{multline}
where for a spin-zero field $h_L = h_R$ and the Euclidean Matsubara frequencies $\omega_{L/R}$ are related to the Lorentzian frequencies $\tilde{\omega}_{L/R}$ by an analytic continuation. The Lorentzian Green's function $G_R$ is given by,
\be
G_R(i \tilde{\omega}_L,i \tilde{\omega}_R) =  G_E (\omega_L,\omega_R).
\ee
We have also included a chemical potential $\Omega_R$ and charge $q_R$ for the right movers for reasons that will soon be clear.

The poles of the Lorentzian retarded correlator lying in the lower half-plane characterize the decay of perturbations in the CFT and are given by the following discrete set of Lorentzian frequencies\footnote{The relation $\Gamma[z] \Gamma[1-z] = \pi \csc (\pi z)$ is helpful to verify the pole structure.}
\begin{eqnarray}
\tilde{\omega}_L  &=&  - 2\pi i T_L (n + h_L), \quad n = 0,1,2,3,\ldots\\
\tilde{\omega}_R  &=&  - 2\pi i T_R (n + h_R) +  q_R \Omega_R, \quad n = 0,1,2,3,\ldots
\end{eqnarray}

The above pole structure can be compared to the poles of the thermal boundary-to-boundary correlator \ref{green} computed earlier. We immediately observe that they have an identical structure provided that we make the following identifications\footnote{Interestingly, the quasinormal modes of the rotating Nariai spacetime obtained in \ref{qnm1} and \ref{qnm2} also have the same structure.}
\begin{eqnarray}
T_L = \frac{1}{2\pi k}, \quad T_R = \frac{\lambda}{4\pi}, \quad \tilde{\omega}_L = m, \quad \tilde{\omega}_R = \tilde{\omega}, \\
\quad h_{L} = h_R = h_{\pm}, \quad q_R = -m, \quad \Omega_R =  \frac{k \lambda}{2}.
\end{eqnarray}

One can recognize $T_L$ as the left-moving temperature used in \ref{cardy}. The right moving temperature $T_R$ is precisely the cosmological temperature observed by observers in the rotating Nariai geometry with non-zero $\lambda$, equation \ref{trn}. The left and right moving frequencies are given by the $\partial_{\hat{t}}$ and $\partial_\phi$ eigenvalues of the scalar modes, and the right moving $U(1)$ charge is also given by the $\partial_\phi$ eigenvalue. Thus, if we are to identify the right moving frequency in the CFT with the $\partial_{\hat{t}}$ eigenvalue we must also posit the existence of a $U(1)$ current algebra whose zero mode coincides with the zero mode of the right moving Virasoro algebra. This is a similar situation to that encountered in the Kerr/CFT correspondence \cite{Bredberg:2009pv,Chen:2010ni}.

\section{Summary and Outlook}

We have explored various aspects of rotating black holes in de Sitter space. Starting with the thermal phase structure, we have discussed a one-parameter family of black hole configurations which has both cosmological and black hole horizons with equal temperature and angular velocity - the rotating Nariai configurations. Geometrically, these solutions are the near horizon region between the black hole and cosmological horizons in the limit where the two coincide. They are given by an $S^2$ fibration over $dS_2$. It is found that they are in an unstable thermodynamic equilibrium, in that small thermal fluctuations result in the system thermally evolving to the most entropic configuration - pure de Sitter space.

Nevertheless, the rotating Nariai geometries serve as a natural starting point for thermal evolution as they can be created from a Euclidean instanton out of nothing \cite{Booth:1998gf,Bousso:1995cc}. Furthermore, they are an interesting type of extremal geometry with a rich asymptotic symmetry group consisting of (at least) one copy of the Virasoro algebra, indicating a possible holographic interpretation \cite{Anninos:2009yc}. Thus, we ventured into the study of scalar perturbations about this geometry.

We uncovered the explicit quasinormal mode structure of this spacetime, as well as the absorption cross-section of the black hole horizon due to an incoming wave originating near the cosmological horizon. Generally, there is a regime in the frequencies of the incoming waves where the absorption cross-section is negative, indicating superradiant scattering. However, in the strict limit where the black hole and cosmological horizons coincide we have found that the absorption cross-section tends to unity and thus superradiance is no longer present. Quantum mechanically, this may imply that spontaneous emission is suppressed in this limit.

Furthermore, we have explored the cosmological properties of the rotating Nariai geometry. We have found evidence for at least three vacua. The $|\text{in}\rangle$ and $|\text{out}\rangle$ vacuum states are those with no incoming particles from $\mathcal{I}^-$ and no outgoing particles at $\mathcal{I}^+$ respectively. Starting at $\mathcal{I}^-$ with the $|\text{in}\rangle$ vacuum, we observed the cosmological production of particles at $\mathcal{I}^+$ and explicitly computed the expectation value of the number of particles produced. We have also proposed that the Euclidean vacuum is simply given by those modes which are analytic in the lower hemisphere of the Euclidean $dS_2$, i.e. $S^2$, part of the geometry. It would be very interesting to put the proposed Euclidean vacuum on a firm footing by carefully examining its analytic structure. Furthermore, it would be extremely interesting to examine the possibility of a complex parameter worth of vacua analogous to the $\alpha$-vacua of de Sitter space \cite{Chernikov:1968zm,Bousso:2001mw} in the rotating Nariai geometry.

Finally, we have computed the boundary-to-boundary correlation functions in the static patch coordinates of the rotating Nariai geometry. The poles of these correlators precisely match the poles of the correlators of a two-dimensional conformal field theory, provided we make a suitable identification of the quantum numbers of the scalar field with those of the operator dual to the scalar in the CFT. This resonates well with the aforementioned proposal that these geometries have a holographic dual given by a two-dimensional conformal field theory. Natural objects to study along this direction would be three-point functions and boundary-to-boundary correlators of vector fields and fermions. The study of fermions in this background might also be motivated by recent results uncovering a Fermi surface in the $AdS_2\times S^2$ near horizon region of an extremal charge black hole in $AdS_4$ \cite{Liu:2009dm,Faulkner:2009wj}. A possible de Sitter analogue might be a Fermi surface in the $dS_2 \times S^2$ near horizon region of the non-rotating Nariai geometry and rotational generalizations thereof.

\section*{Acknowledgements}

It has been a great pleasure discussing this work with J. Barandes, M. Esole, F. Denef and A. Strominger. We also thank H. Cholula, H. Disch and C. Emil for their inspiration. D.A. was supported in part by
DOE grant DE-FG02-91ER40654.

\appendix

\section{Explicit Thermodynamic Expressions}\label{expressions}

Below we present explicit expressions for the specific heat at fixed angular momentum, $C_J$, the second partial derivative of the entropy with respect to the angular momentum and the determinant of the Hessian given in \ref{conditions}. Our results are given in the $(r_+,a)$-basis.
\newline
\newline
The specific heat is given by:
\begin{multline}\label{specheat}
C_J(r_+,a) = \frac{1}{\left(a^2+\ell^2\right)} \times \\ \frac{2 \ell^4 \pi  \left(a^2+r_+^2\right){}^2 \left(3 r_+^4-\ell^2 r_+^2+a^2 \left(\ell^2+r_+^2\right)\right)}{
   \left(\ell^2-r_+^2\right) a^6+\left(-3 \ell^4+13 r_+^2 \ell^2-6 r_+^4\right) a^4+r_+^2 \left(-6 \ell^4+23 r_+^2 \ell^2-9 r_+^4\right)
   a^2+\ell^2 r_+^4 \left(\ell^2+3 r_+^2\right)}.
\end{multline}
The second derivative of the entropy with respect to $J$ is given by:
\begin{multline}
\frac{\partial^2 S_{BH}}{\partial J^2} = 8 \left(a^2+\ell^2\right) \pi  r^2 \times \\ \frac{\left(-3 \ell^4+6 r^2 \ell^2+r^4\right) a^4+\left(\ell^6-13 r^2 \ell^4+23 r^4 \ell^2-3 r^6\right) a^2+\ell^2
   r^2 \left(\ell^2-3 r^2\right)^2}{\ell^2 \left(3 r^4-\ell^2 r^2+a^2 \left(\ell^2+r^2\right)\right)^3}.
\end{multline}
The determinant of the Hessian is given by:
\be\label{hessian}
\det H_{ij} = -\frac{64 \left(a^2+\ell^2\right)^4 \pi ^2 r_+^4 \left(\left(\ell^2-r_+^2\right) a^2+r_+^2 \left(\ell^2+3 r_+^2\right)\right)}{\ell^2 \left(3
   r_+^4-\ell^2 r_+^2+a^2 \left(\ell^2+r_+^2\right)\right){}^4}.
\ee
We note that it is negative definite in the physical configuration space.

\section{Thermal Evolution}

We obtain three qualitative possibilities for thermal evolution assuming that the emission rates are dominated by angular momentum, energy and a combination thereof. Below, we show plots of the vector field $(c_J \; \partial_J S_{tot},c_M \; \partial_M S_{tot})$ where $c_J$ and $c_M$ are constants determining the relative emission rate.
\newline
\newline
The first case corresponds to a situation where there is an `energy pump' between the two horizons keeping the energy fixed for each horizon, i.e. $c_M \ll 1$. In this case, depicted in Fig. 5 (a), the thermal flow will be along lines of constant energy depicted in Fig. 3 (b), and will lead to the complete spin-down of the black hole.
\begin{figure}[!h]\label{thevol}
\begin{center}
$\begin{array}{c@{\hspace{0.1in}}c}
\includegraphics[angle=0,width=0.5\textwidth]{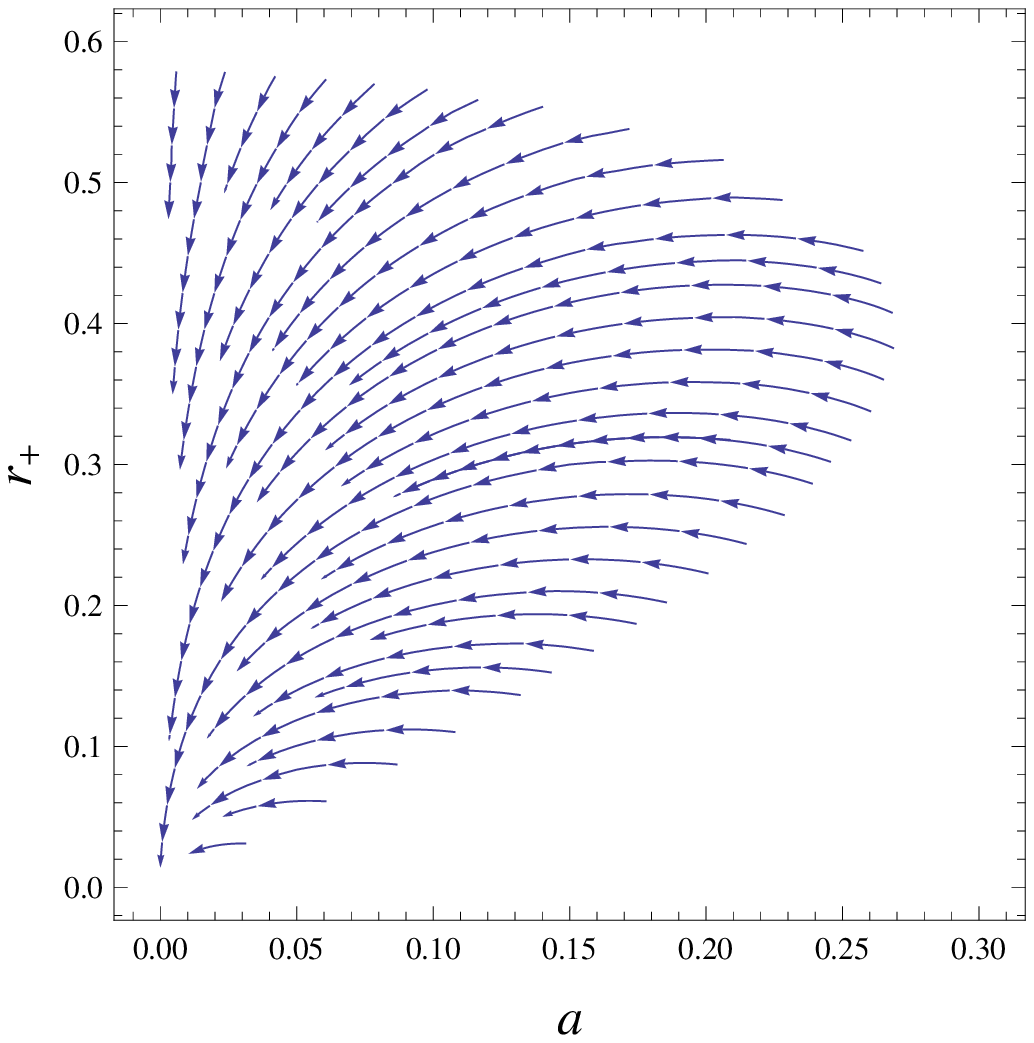} &
 \includegraphics[angle=0,width=0.5\textwidth]{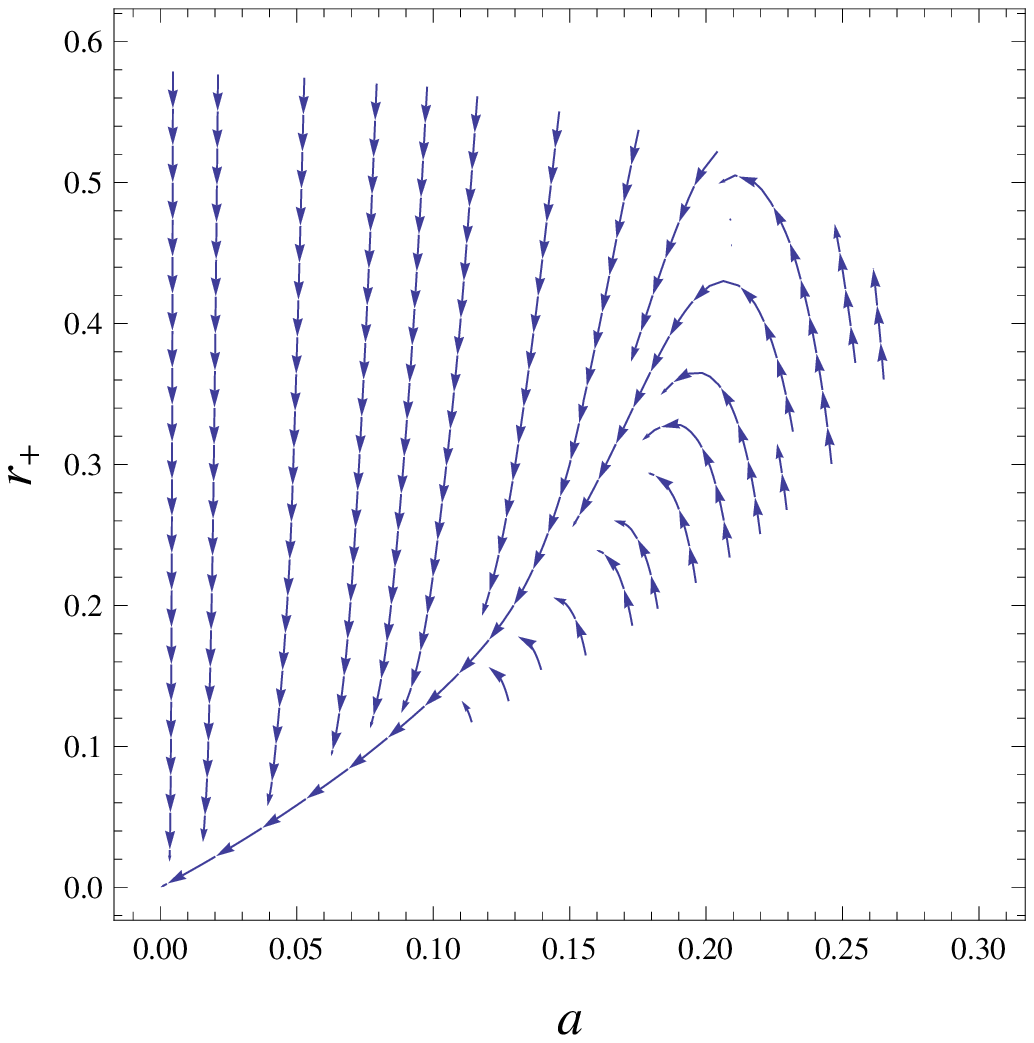} \\
   \text{(a)}&\text{(b)}
   \end{array}$
\end{center}
\caption{(a): Thermal evolution when emission of energy is suppressed. (b): Thermal evolution when emission of angular momentum is suppressed. }
\end{figure}
\newline
\newline
The second case corresponds to a situation where there is an `angular momentum' pump between the two horizons keeping the angular momentum fixed for each horizon and depicted in Fig. 5 (b), i.e. $c_J \ll 1$. One notices that large extremal black holes evolve toward the rotating Nariai limit before spinning down and evolving toward the origin. Notice that once the lukewarm line is reached the system evolves along it.
\begin{figure}[!t]\label{}
\begin{center}
\includegraphics[angle=0,width=0.5\textwidth]{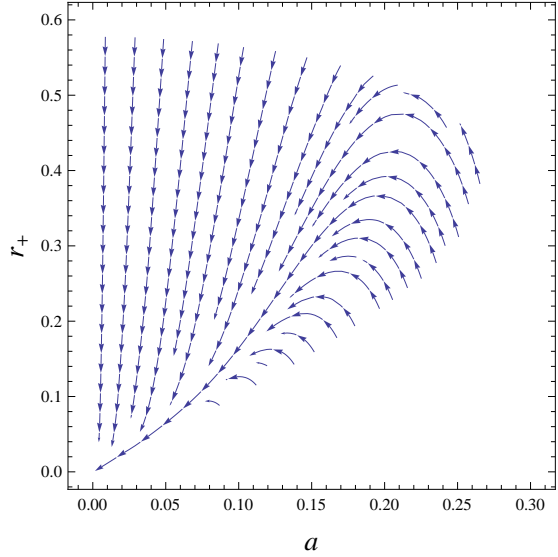}
\end{center}
\caption{ Thermal evolution when emission of energy and angular momentum is comparable. }
\end{figure}
\newline
\newline
Finally, we can consider a situation where both angular momentum and energy are emitted at similar rates, i.e. $c_J \sim c_M$, as depicted in Fig. 6.

\section{Explicit Expressions for $\alpha$ and $\beta$}\label{alphabeta}

The explicit expressions for $\alpha$ and $\beta$ are as follows:
\begin{multline}
\alpha = \frac{\lambda ^{h_--i(n_R+\tilde{\omega}/\lambda)} \csc \left[\pi  \left(h_--i(n_R+\tilde{\omega}/\lambda\right)\right] \Gamma [h_--i k
   m]  \Gamma \left[h_-+i \left(n_R+\tilde{\omega}/\lambda\right)\right]}{\pi\Gamma [2 h_-]Q	},
\end{multline}

\begin{multline}
\beta = -\frac{\lambda ^{h_+-i(n_R+\tilde{\omega}/\lambda)} \csc \left[\pi  \left(h_-+i(n_R+\tilde{\omega}/\lambda\right)\right] \Gamma [h_+-i k
   m]  \Gamma \left[h_++i \left(n_R+\tilde{\omega}/\lambda\right)\right]}{\pi\Gamma [2 h_+]Q	}
\end{multline}
where we have defined:
\begin{multline}
\frac{Q}{\Gamma \left[1-\frac{2 i \tilde{\omega} }{\lambda }\right]]} \equiv \csc [\pi (h_--i k m)] \csc\left[
   \pi  \left(h_--i (n_R +\tilde{\omega}/\lambda )\right)\right]\\
-\csc [\pi(h_-+i k m)] \csc \left[\pi  \left(h_-+i \left(n_R+\tilde{\omega}/\lambda\right)\right)\right].
\end{multline}

\section{Thermal Boundary-to-Boundary Correlator}\label{twopoint}

We will discuss in some more detail the computation of the retarded thermal boundary-to-boundary correlator.\footnote{A similar situation for $AdS$ is discussed in \cite{Klebanov:1999tb,Son:2002sd}.} Begin with a wave-packet expressed as a sum over modes
\be
\Phi(x,\hat{t},\Omega) = \sum_{m,l} \int \frac{d\tilde{\omega}}{2\pi} \left( \gamma_{l m \tilde{\omega}} e^{- i\tilde{\omega} \hat{t}} e^{i m \hat{\phi}} Y_{l m}(\theta) R(x) \right).
\ee
The on-shell action for a massless scalar field is given by
\be
I_{matter} = \frac{1}{2} \int_{\mathcal{M}} d^4 x \sqrt{-g} \partial_\mu \Phi \partial^\mu \Phi =  \frac{1}{2}\int_{\partial\mathcal{M}} d^3x \sqrt{-g} n^\mu \Phi \partial_\mu \Phi
\ee
where we have integrated by parts and set the bulk integral to zero since it vanishes on-shell. The only boundary term relevant to us will be the one at $\mathcal{I}^+$. The $n^\mu$ is a unit normal vector which is orthogonal to the boundary. Using the late time behavior of the $R(x)$ function in the rotating Nariai limit \ref{rtot} and evaluating the action we find
\be
I_{matter} = \frac{1}{2} \sum_{m, l} \int \frac{d\tilde{\omega}}{2\pi} \alpha(\tilde{\omega}, m, l) \beta(-\tilde{\omega}, -m, l) + \ldots
\ee
where the dots correspond to terms that oscillate infinitely fast at the boundary. We may drop such terms by adding a small imaginary part to $x$, as is done for the vacuum state of the harmonic oscillator.

In order to obtain this expression, we have used the following completeness and orthonormality properties of the spheroidal harmonics,
\begin{eqnarray}
\sum_{l,m} Y_{l m}(\theta) e^{i m \hat{\phi} }Y_{l m} (\theta') e^{-i m \hat{\phi}'} &=& \delta^2(\Omega, \Omega'), \\
\int d\theta d\hat{\phi} \sqrt{\tilde{h}} Y_{l m}(\theta) e^{i m \hat{\phi} }Y_{l' m'} (\theta) e^{-i m' \hat{\phi}} &=& \delta_{m-m'}\delta_{l - l'}
\end{eqnarray}
where $\tilde{h}_{ij}$ is the induced metric at fixed $x$ and $t$ and we are working in a basis where $Y_{l m}(\theta)$ are real.

To compute the thermal boundary-to-boundary correlator, we must define which excitation, i.e. $\alpha$ or $\beta$, corresponds to the boundary value of the field. Once we have defined the boundary value, we can take variational derivatives with respect to it. For instance, choosing $\alpha$ as the boundary value at $\mathcal{I}^+$ we can write the action as
\be
I_{matter} = \frac{1}{2} \sum_{m, l} \int \frac{d\tilde{\omega}}{2\pi} \alpha(\tilde{\omega}, m, l) \alpha(-\tilde{\omega}, -m, l) G^{th}_R(l,m,\tilde{\omega})
\ee
where the thermal boundary-to-boundary correlator in momentum space is thus found to be:
\be
G^{th}_R \sim \frac{\beta}{\alpha}.
\ee

\end{document}